\documentclass[epj,referee]{svjour}
\usepackage{graphics}
\begin{document}
\title{Aperiodicity-induced effects on the transmission resonances in
       multibarrier systems}

\author{ Gi-Yeong Oh \thanks{e-mail: ogy@anu.ansung.ac.kr} }

\institute{Department of Basic Science, Hankyong National University,
           Kyonggi-do 456-749, Korea}

\date{Received: \today}

\abstract{
We study the resonant tunneling properties of an electron through a few
types of binary periodic and aperiodic multibarrier systems. Within the
framework of the effective-mass approximation, we calculate the
transmission coefficients to investigate the dependence of the
transmission resonances on the system parameters such as the kind of
aperiodicity, the generation number, and the widths of the wells and
barriers. Similarities and differences of the resonances between the
binary periodic and aperiodic systems are discussed in detail.
Transmission resonances in aperiodic systems are found to be
characterized by complex resonance splitting and a variety of
peak-to-valley ratios which are not exhibited in the periodic system.
For some energy ranges, transmission resonances in aperiodic systems
are also found to resemble those in the periodic system, despite the
existence of aperiodicity.
\PACS{
   {73.40.Gk}{Tunneling} \and
   {71.55.Jv}{Disordered structures; amorphous and glassy solids} \and
   {73.20.Dx}{Electron states in low-dimensional structures}
     }
}

\maketitle

\section{Introduction}

Tunneling of an electron through a potential barrier is one of the
fundamental phenomena in quantum mechanics and plays a key role in the
physics of electronic and optoelectronic devices [1]. Stimulated by the
advancement in modern material fabrication technology such as
molecular-beam epitaxy and metal-organic chemical vapor deposition,
there has been a lot of work on the problem of the electronic resonant
tunneling in semiconductor superlattices [2-15]. Among interesting
features emerged from the studies, the most well recognized features
are the resonance splitting effects and the energy band effects; for a
finite superlattice composed of $N$ identical potential barriers with
arbitrary profiles at zero bias, there occurs $(N-1)$-fold resonance
splitting, and the split resonant energies approach the band structure
for large $N$ [1, 12-14]. Very recently, Guo {\it et al.} [15] studied
the resonance splitting effects in superlattices which are periodically
juxtaposed with two different building barriers to demonstrate that the
resonance splitting is determined not only by the structure but also
by the parameters of building barriers.

In a different context, there has been much interest in the electronic
properties of deterministic aperiodic systems [16-19] which are known
to have more complex geometrical structures than the periodic system.
Studies on these systems have revealed a variety of exotic electronic
properties such as the singular continuity of the energy spectrum, the
self-similarity of the electronic wave functions, the power-law
behavior of the resistance, and so forth. However, most of the work has
focused on the electronic properties in the infinite limit of the
system size [16-20], and the study on the transport properties of the
finite-size systems, particularly from tunneling point of view, has
received less attention. Recently, Singh {\it et al.} [21] studied the
electronic transport properties of the Fibonacci and Thue-Morse (TM)
superlattices to compare the results with those of the periodic system.
Besides, Liu {\it et al.} [22] calculated the electronic transmission
spectra of the Cantor fractal multibarrier systems to show that the
tunneling spectrum is more complex than that of the periodic system.

\begin{figure}
\centerline{
\resizebox{0.40\textwidth}{!}{ \includegraphics{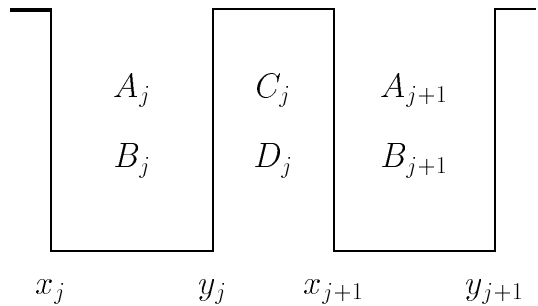} }
}
\caption{Schematic configuration of a part of the multibarrier systems
 used in the calculation.}
\label{fig:1}
\end{figure}

In this paper, we study the aperiodicity-induced effects on the
transmission resonances in a few types of binary aperiodic multibarrier
systems. To do this, taking into account four kinds of binary
multibarrier systems whose geometrical structures are determined by
Eq.~(11), we calculate the transmission coefficients of an electron
through these multibarrier systems. Dependence of the transmission
resonances on the system parameters such as the kind of aperiodicity,
the generation number, and the widths of the wells and barriers is
investigated. From this, we first illustrate the characteristics of the
transmission resonances exhibited in the binary periodic (BP) system,
and then make a comparison of the resonances between the BP and
aperiodic systems; similarities and differences between them are
presented in detail. In doing this, a comparison of the resonance
splitting exhibited in the common-well (CW) structure with those
exhibited in the common-barrier (CB) structure of the systems is also
presented.

\section{Method}

We shall now derive an expression for the transmission coefficient of a
multibarrier system using the transfer matrix formalism. To do this, we
consider an electron with a longitudinal energy $E$ incident from left
of the system. Assuming that the phonon scattering can be neglected and
no bias is applied across the system, the effective-mass approximation
leads to the continuous Schr\"{o}dinger equation
\begin{equation}
 \left[-\frac{\hbar^{2}}{2m_{j}^{*}}\frac{d^{2}}{dx^{2}}
 +V(x)\right]\Psi_{j}(x)=E\Psi_{j}(x) ,
\end{equation}
where $x$ is the longitudinal direction of the system, $V(x)$ the
minimum energy of the conduction band, and $m_{j}^{*}=m_{w(b)}^{*}$ the
effective mass of the electron in the well (barrier) of the $j$th cell.
Figure~1 shows the schematic configuration of a part of the system. For
convenience of calculation, we set the conduction band minimum to be
zero and the potential barrier to be rectangular, i.e.,
\begin{equation}
 V(x)= \left\{\begin{array}{lcl}
 0 &~~{\rm for}~~& x_{j}<x<y_{j}\\ V &~~{\rm for}~~& y_{j}<x<x_{j+1}
 \end{array} \right.,
\end{equation}
where $x_{j}$ $(y_{j})$ is the starting position of the $j$th well
(barrier). Then, the wave function associated with the electron in the
$j$th cell can be written as
\begin{equation}
 \psi_{j}(x)=A_{j}e^{i k (x-x_{j})}+B_{j}e^{-ik(x-x_{j})}
\end{equation}
in the well and
\begin{equation}
 \phi_{j}(x)=C_{j}e^{-\kappa(x-y_{j})}+D_{j}e^{\kappa(x-y_{j})}
\end{equation}
in the barrier. Here,
\begin{equation}
 k=\sqrt{\frac{2m_{w}^{*}E}{\hbar^{2}}},~~~
 \kappa=\sqrt{\frac{2m_{b}^{*}(V-E)}{\hbar^{2}}}
\end{equation}
are the wave numbers in the wells and barriers, respectively.

By applying the Bastard's matching conditions of the wave function and
its derivative at discontinuity of $V(x)$, we can write the relation of
the coefficients between the $j$th and $(j+1)$th wells as
\begin{equation}
 \left(\begin{array}{c} A_{j+1} \\ B_{j+1} \end{array}\right)
 =T_{j}\left(\begin{array}{c} A_{j}\\ B_{j} \end{array} \right)
 =\left(\begin{array}{cc} \alpha_{j} & \beta_{j} \\
 \beta_{j}^{*} & \alpha_{j}^{*} \end{array}\right)
 \left(\begin{array}{c} A_{j}\\ B_{j} \end{array} \right),
\end{equation}
where $T_{j}$ is the unimodular transfer matrix, and $\alpha_{j}$ and
$\beta_{j}$ are given by
\begin{eqnarray}
 \alpha_{j}&=&\left[\cosh(\kappa b_{j})
 +\frac{i}{2}\left(\frac{m_{b}^{*}k}{m_{w}^{*}\kappa}
 -\frac{m_{w}^{*}\kappa}{m_{b}^{*}k}\right)\sinh(\kappa b_{j})
 \right]e^{ikw_{j}}, \nonumber\\
 \beta_{j}&=&-\frac{i}{2}\left(\frac{m_{b}^{*}k}{m_{w}^{*}\kappa}
 +\frac{m_{w}^{*}\kappa}{m_{b}^{*}k} \right)\sinh(\kappa
 b_{j})e^{-ikw_{j}}.
\end{eqnarray}
Here, $w_{j} (=y_{j}-x_{j})$ and $b_{j} (=x_{j+1}-y_{j})$ are the
widths of the $j$th well and barrier. Multiplying $T_{j}$ successively,
we can write the relation of $A$'s and $B$'s between the first and the
$(N+1)$th region as
\begin{equation}
 \left(\begin{array}{c} A_{N+1} \\ B_{N+1} \end{array}\right)
 =M_{N}\left(\begin{array}{c} A_{1}\\ B_{1} \end{array}\right),
\end{equation}
where $M_{N}$ is the total transfer matrix given by
\begin{equation}
 M_{N}=T_{N}T_{N-1}\cdots T_{2}T_{1}
 =\left(\begin{array}{cc} a_{N} & b_{N} \\ b_{N}^{*} & a_{N}^{*}
 \end{array}\right).
\end{equation}
Since there will be a reflected wave in the first region but only a
transmitted wave in the $(N+1)$th region (i.e., $B_{N+1}=0$), we can
write the transmission coefficient as
\begin{equation}
 T=\frac{1}{|a_{N}^{*}|^{2}}.
\end{equation}
Generally, it requires extensive matrix manipulation to calculate $T$.
However, for the multibarrier systems considered in this paper, $T$ can
be easily calculated in terms of the deterministic substitution rules
given below.

We now introduce four kinds of deterministic sequences $-$ the BP, the
TM [16,17], the period-doubling (PD) [18], and the
copper-mean (CM) [19] sequences which are generated by the substitution
rules
\begin{equation}
  \begin{array}{lll}
    BP :& S_{l+1}=S_{l}^{2},& S_{1}=AB \\
    TM :& S_{l+1}=S_{l}\overline{S}_{l},& (\overline{S}_{0},S_{0})
    =(B,A)\\
    PD :& S_{l+1}=S_{l}S_{l-1}^{2},& (S_{0},S_{1})=(A,AB) \\
    CM :& S_{l+1}=S_{l}S_{l-1}^{2},& (S_{-1},S_{0})=(B,A),
  \end{array}
\end{equation}
where $l$ is the generation number (i.e., $N=2^{l}$ for the first three
sequences and $N=[2^{l+2}-(-1)^{l}]/3$ for the last sequence) and
$\overline{S}_{l}$ is the complement of $S_{l}$ which is obtained by
exchanging the letters $A$ and $B$. Here, $A$ and $B$ represent two
kinds of unit cells of the multibarrier system. The unit cell $A$ ($B$)
consists of a well with the width $w_{A}$ ($w_{B}$) and a barrier with
the width $b_{A}$ ($b_{B}$) and the height $V$. We refer the case of
$w_{A}=w_{B}$ with $b_{A}\neq b_{B}$ as the CW model and the case of
$b_{A}=b_{B}$ with $w_{A}\neq w_{B}$ as the CB model, respectively.

By means of Eq.~(11), we can write the recursion relations of the total
transfer matrices between different generations as
\begin{equation}
  \begin{array}{lll}
    BP:& M_{l+1}=M_{l}^{2},& M_{1}=T_{B}T_{A} \\
    TM:& M_{l+1}=\overline{M}_{l}M_{l},& (\overline{M}_{0},M_{0})
    =(T_{B},T_{A})\\
    PD:& M_{l+1}=M_{l-1}^{2}M_{l},& (M_{0},M_{1})=(T_{A},T_{B}T_{A})\\
    CM:& M_{l+1}=M_{l-1}^{2}M_{l},& (M_{-1},M_{0})=(T_{B},T_{A}) .
  \end{array}
\end{equation}
Using the relations in Eq.~(12), we can easily derive the recursion
relations of $a$'s and $b$'s between different
generations as follows:
\begin{equation}
 \left\{\begin{array}{l}a_{l+1}=a_{l}^{2}+b_{l}^{*}b_{l} \\
  b_{l+1}=b_{l}(a_{l}+a_{l}^{*}) \end{array} \right.
\end{equation}
with $a_{1}=\alpha_{A}\alpha_{B}+\beta_{A}^{*}\beta_{B}$ and
$b_{1}=\alpha_{B}\beta_{A}+\alpha_{A}^{*}\beta_{B}$ for the BP
sequence,
\begin{equation}
 \left\{\begin{array}{ll}
  a_{l+1}=\overline{a}_{l}a_{l}+\overline{b}_{l}b_{l}^{*},~&
  b_{l+1}=\overline{a}_{l}b_{l}+a_{l}^{*}\overline{b}_{l} \\
  \overline{a}_{l+1}=a_{l}\overline{a}_{l}+b_{l}\overline{b}_{l}^{*}
  ,~& \overline{b}_{l+1}=a_{l}\overline{b}_{l}+
  \overline{a}_{l}^{*}b_{l} \end{array} \right.
\end{equation}
with $a_{0}=\alpha_{A}$, $b_{0}=\beta_{A}$,
$\overline{a}_{0}=\alpha_{B}$, and $\overline{b}_{0}=\beta_{B}$ for the
TM sequence, and
\begin{equation}
 \left\{\begin{array}{l}a_{l+1}=a_{l}(a_{l-1}^{2}+|b_{l-1}|^{2})
 +b_{l}^{*}b_{l-1}(a_{l-1}+a_{l-1}^{*}) \\
 b_{l+1}=b_{l}(a_{l-1}^{2}+|b_{l-1}|^{2})
 +a_{l}^{*}b_{l-1}(a_{l-1}+a_{l-1}^{*}) \end{array} \right.
\end{equation}
with $a_{0}=\alpha_{A}$, $b_{0}=\beta_{A}$,
$a_{1}=\alpha_{A}\alpha_{B}+\beta_{A}^{*}\beta_{B}$, and
$b_{1}=\alpha_{B}\beta_{A}+\alpha_{A}^{*}\beta_{B}$ for the PD
sequence. Recursion relations for the CM sequence are exactly the same
as Eq.~(15) with $a_{-1}=\alpha_{B}$, $b_{-1}=\beta_{B}$,
$a_{0}=\alpha_{A}$, and $b_{0}=\beta_{A}$.

\section{Numerical results and discussion}

As a sample material for calculation, we choose the
$\mu$c-Si:H/$a$-Si:H superlattice [22], where $\mu$c-Si:H acts as the
well and $a$-Si:H the barrier with $V=0.4$ eV. The effective mass in
the wells and barriers is taken to be $m_{w}^{*}=m_{b}^{*}=0.3m_{e}$
[23], where $m_{e}$ is the free electron mass. In calculating
transmission coefficients of an electron through multibarrier systems,
we treat two cases separately; the one is to set the widths of the
wells equal while arrange the widths of the barriers according to the
given substitution rule (the CW model), and the other is to set the
widths of the barriers equal while arrange the widths of the wells
according to the given substitution rule (the CB model). Some examples
of the results are plotted in Figures~2 and 3. In plotting, the mesh of
$E$ is taken to be $\Delta E=1.0\times 10^{-5}$ eV, and the system
parameters are set to be $w_{A}=w_{B}=20$ \AA~ and $b_{A}=2b_{B}=8$
\AA~ in the CW model, and $b_{A}=b_{B}=7$ \AA~ and $w_{A}=2w_{B}=40$
\AA~ in the CB model, respectively. Figure~2 shows the results obtained
from the CW model for the energy range near the lowest domain of
resonances, and Figure~3 shows the results obtained from the CB model
for the energy range near the first two lowest domains of resonances.
Here, the `domain of resonances' means the energy range that contains
resonant peaks in the finite-size system and would approach the allowed
energy band in the infinite limit of the system size.

We first discuss the features of the transmission resonances exhibited
in the BP system with $l=2$. In this case, three resonant peaks in each
domain of resonances are expected due to overlap of quasi-bound states
in the three well regions, and the result obtained from the CW model
[Figure~2a] agrees well with the expectation. An interesting feature to
be noted in Figure~2a is that the first and the third peaks of the
three resonant peaks are complete (i.e., $T=1$) while the second peak
is incomplete (i.e., $T<1$). We will see that the two complete peaks
locate in the middle of the subdomains while the incomplete peak
disappears for large $l$ [see Figure~2e]. As for the result obtained
from the CB model [Figure~3a], there are two distinctive features from
the result obtained from the CW model. The one is that there occurs
suppression of resonant peaks. We can see in Figure~3a that there exist
a single complete peak in the first lowest domain and two complete
peaks in the second lowest domain, which implies that two peaks in the
first lowest domain and one peak in the second lowest domain are
suppressed. The second is that the resonance widths exhibited in the CB
model are much narrower and sharper than those in the CW model.

\begin{figure}
\centerline{
\resizebox{0.40\textwidth}{!}{ \includegraphics{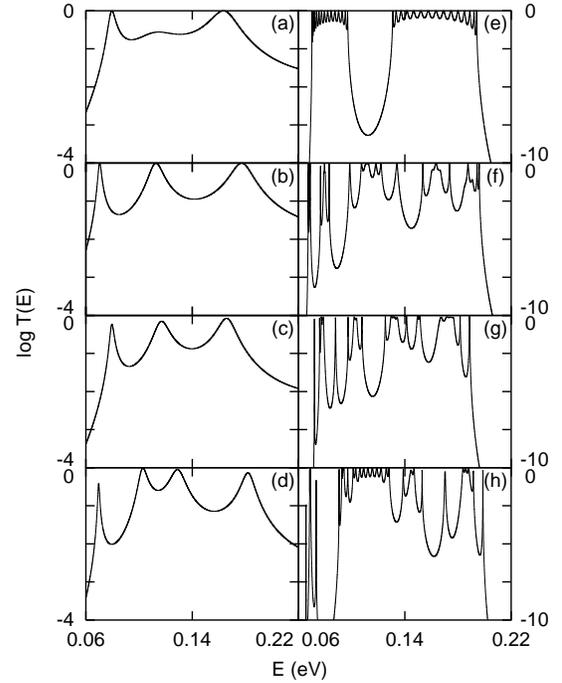} }
}
\caption{$\log_{10}T(E)$ versus $E$ near the
lowest domain of resonances for the BP [(a) and (e)], TM [(b) and (f)],
PD [(c) and (g)], and CM [(d) and (h)] systems with
$w_{A}=w_{B}=20$ \AA~ and $b_{A}=2b_{B}=8$ \AA~ (the CW model). The
generation number is $l=2$ for (a)-(d) and $l=5$ for (e)-(h).}
\label{fig:2}
\end{figure}

\begin{figure}
\centerline{
\resizebox{0.40\textwidth}{!}{ \includegraphics{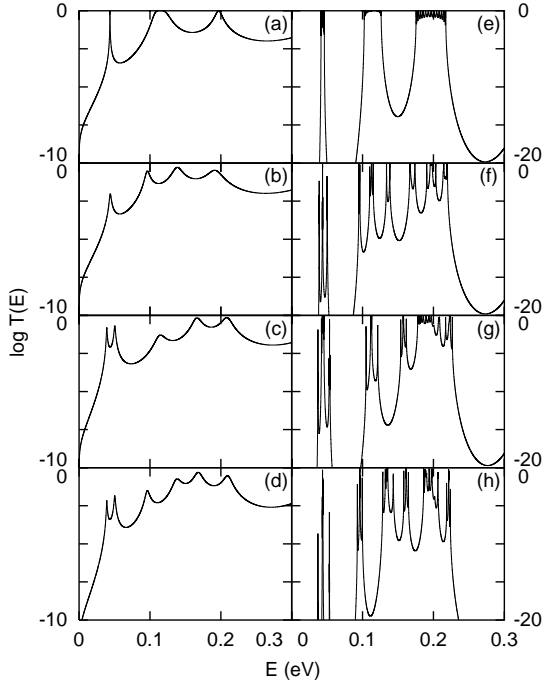} }
}
\caption{$\log_{10}T(E)$ versus $E$ near the
first two domains of resonances for the BP [(a) and (e)], TM [(b) and
(f)], PD [(c) and (g)], and CM [(d) and (h)] systems with
$w_{A}=2w_{B}=40$ \AA~ and $b_{A}=b_{B}=7$ \AA~ (the CB model). The
generation number is $l=2$ for (a)-(d) and $l=5$ for (e)-(h).}
\label{fig:3}
\end{figure}

We now discuss the features of the transmission resonances in the BP
system with large $l$. As $l$ increases, successive resonant splitting
effects and the energy band effects on the transmission properties are
expected, and we confirm them. Figure~2e shows the transmission
coefficients obtained from the CW model with $l=5$. Here two
distinctive features from the case of the periodic system [12-14] are
emerged. The first is that the main domain splits into two subdomains,
the centers of which correspond to the first and the third resonant
peaks in Figure~2a. We argue that the occurrence of this kind of
resonance splitting attributes to the binary periodicity of the system.
The second is that the total number of resonant peaks is not exactly
the same as the number of the wells. In Figure~2e, each subdomain
contains 15 peaks and the total number of peaks are 30 while there are
31 wells in the system with $l=5$. The result obtained from the CB
model with $l=5$ is plotted in Figure~3e. Similar features appeared in
the CW model are also exhibited in the second lowest main domain; the
domain splits into two subdomains and the total number of resonant
peaks is less than the number of the wells. However, the transmission
coefficients in the lowest main domain display somewhat different
behavior; no splitting into subdomains occurs, which indicates that the
binary periodicity of the system does not affect on the splitting of
this domain.

We also study the features of the transmission resonances with varying
the widths of the wells and barriers, and observe that the number of
resonance domains increases with increasing the width of the well,
which can be easily understood by noting that the energies of the
quasi-bound states in a well are approximately in proportional to the
inverse square of the width of the well [24]. We also observe that the
widths of resonance domains decrease and the peak-to-valley ratios
increase with increasing the width of the barrier.

Having seen the features of the transmission resonances in the BP
system, we now discuss the features of the transmission resonances
exhibited in the aperiodic TM, PD, CM systems with $l=2$. In this case,
the three, three, and four resonant peaks are expected to exist in each
domain of resonances due to overlap of quasi-bound states in the three,
three, and four wells of the TM, PD, CM systems. The results obtained
from the CW model for these systems are plotted in Figures~2b $-$ 2d,
where it can be seen that the number of resonant peaks fits with the
expectation. However, the results obtained from the CB model
[Figures~3b $-$ 3d] do not always fit with the expectation: The
transmission coefficients for the second lowest domain agree with the
expectation, while the number of resonant peaks in the first lowest
domain is less than the expected number due to suppression of the
peaks. The number of suppressed peaks is two, one, and one in the TM,
PD, CM system, respectively. As for the effects of aperiodicity, we
would like to mention two points. The first is that the lowest resonant
peaks shift towards the lower energy region, compared with that of the
BP system. The second is that there coexist the complete and incomplete
resonant peaks. For the peaks exhibited in the CW model, all the peaks
in the TM system and the second peak in the CM system are complete,
while all the peaks in the PD system and all the peaks expect the
second peak in the CM system are incomplete. In the meanwhile, for the
peaks exhibited in the CB model, all the peaks are incomplete, which
implies that it is more difficult for an electron to tunnel through the
CB structure than the CW structure. Here it should be noted that the
incomplete resonant peak appeared in the BP system [Figure~2a] and the
incomplete peaks appeared in the aperiodic systems show different
behavior as $l$ increases; the former fails to locate in the middle of
the subbands and disappears, while the latter survives to locate in the
middle of the subdomains

We now discuss the features of the transmission resonances in the
aperiodic systems with large $l$. Figures~2f $-$ 2h show the results
obtained from the CW model of the TM, PD, and CM systems with $l=5$,
which reveal two distinctive features to be mentioned. The first is
that, even though the aperiodicity of each system is binary as in the
BP system, the resonance splitting pattern is much more complex than
that of the BP system such that a variety of peak-to-valley ratios
appear. We argue that the level of hierarchy in the resonance splitting
will become deeper with the increase of $l$ and the split resonant
peaks will eventually compose of Cantor-like fractal structure in the
infinite limit of the system size. The second is that, despite the
existence of aperiodicity, the transmission resonances for some energy
ranges resemble those exhibited in the BP system. For example, from
$0.101$ eV to $0.128$ eV and from $0.184$ eV to $0.188$ eV in
Figure~2f, the transmission coefficients exhibit the `resonance
plateaus' as in the case of the BP system. We also find that, for
higher energy ranges, the effect of aperiodicity weakens such that the
number and the widths of resonant plateaus increase, which resembles
the feature of the periodic system. The results obtained from the CB
model of the TM, PD, and CM systems [Figures~3f $-$ 3h] reveal similar
features to those exhibited in the CW model; complex resonance
splitting effects and the existence of the resonance plateaus are
clearly seen. An example of the resonance plateaus can be seen in
Figure~3g with the energy range from $0.178$ eV to $0.197$ eV.

\section{Summary}

In summary, we studied the effects of aperiodicity on the transmission
resonances of an electron through a few types of binary aperiodic
multibarrier systems with finite system size. Dependence of
transmission resonances on the system parameters was investigated, from
which the similarities and differences of the resonances between the
binary periodic and aperiodic systems were presented. In doing this, a
comparison of resonance splitting exhibited in the common-well
structure with those exhibited in the common-barrier structure of the
multibarrier systems was also made in detail. It was found that complex
resonance splitting and a variety of peak-to-valley ratios which are
not exhibited in the periodic system are emerged as a result of
introducing aperiodicity. It was also found that the transmission
resonances for some energy ranges in the aperiodic systems resemble
those in the periodic system, despite the existence of aperiodicity. We
hope that exotic resonant tunneling properties of the binary aperiodic
multibarrier systems such as the complex resonance splitting effects,
deep levels of hierarchy in the peak-to-valley ratios, and the
existence of tunneling plateaus can be applied in designing a new type
of electronic device.

\end{document}